# Perfect anomalous splitter by acoustic meta-grating

Huiqin Ni[1], Xinsheng Fang[2], Zhilin Hou[1*] Yong Li[2*] and Badreddine Assouar[3*]

[1]School of Physics and Optoelectronics, South China University of Technology, Guangzhou 510640, P. R. China

[2]Institute of Acoustics, School of Physics Science and Engineering, Tongji University. Shanghai 200092, P. R. China

[3]Institut Jean Lamour, Université de Lorraine, CNRS, F-54000 Nancy, France

As an inversely designed artificial device, metasurface usually means densely arranged meta-atoms with complex substructures. In acoustics, those meta-atoms are usually constructed by multi-folded channels or multi-connected cavities of deep sub-wavelength feature, which limits their implementation in pragmatic applications. We propose here a comprehensive concept of a perfect anomalous splitter based on an acoustic meta-grating. The beam splitter is designed by etching only two or four straight-walled grooves per period on a planar hard surface. Different from the recently reported reflectors or splitters, our device can perfectly split an incident wave into different desired directions with arbitrary power flow partition. In addition, because ultrathin substructures with thin walls and narrow channels are avoided in our design procedure, the proposed beam splitter can be used for waves with much shorter wavelength compared to the previous suggested systems. The design is established by rigorous formulae developed under the framework of the grating theory and a genetic optimization algorithm. Numerical simulation and experimental evidence are provided to discuss the involved physical mechanism and to give the proof-of-concept for the proposed perfect anomalous acoustic splitter.

# Introduction

High efficient wave manipulation via artificial structures is always strongly desired in materials physics and engineering communities. The design of a thin material which can control the wave propagation in a desired manner is highly intriguing, yet greatly challenging. Since 2011, the concept of metasurface, namely a two-dimensional (2D) thin artificial material/structure, have been introduced firstly in electromagnetic wave system[1] and extended into acoustic one[2,3]. This burgeoning field of rationally designed 2D materials of sub-wavelength thickness opens a new degree of freedom for sound wave manipulation. They provide unique functionalities with large potential of engineering applications such as anomalous refraction and reflection[4-9], asymmetric transmission[10,11], holograms[12,13], perfect absorptions[14,15], retroreflection [16,17] and cloaking[18].

Acoustic metasurfaces usually contain several kinds of unit cells providing additional $2\pi$ span phase shifts (tangential momentum). Then, a discrete phase profile can be constructed using these individual unit cells by fitting a continuous phase profile derived from a desired pressure field[19,20]. As a result, the wave scattered from a metasurface forms the desired pattern. Because these phase profiles are usually complex and have to consider the resolution in discretization, the implemented unit cells, in general, possess a geometrical size in deep sub-wavelength scale with elaborated configurations. This leads to the inevitably loss and deformation effects inside the unit cells along with a reduced efficiency. Meanwhile, the design strategy implies that the unit cells are individually conceived without considering the energy interchange among them. Under an oblique incidence with a large angle, the coupling effect becomes prominent, and generates unexpected inside lobes (higher orders of diffractions). To overcome this drawback, bi-anisotropic metasurfaces[21,22], which could redirect the normal incident wave to large angles with over 90% efficiency has been suggested. However, they still suffer from the sub-wavelength unit cells and imperfect performance.

On the other hand, based on diffraction grating principles[23,24], the configuration for acoustic anomalous reflection can be greatly simplified as the composed substructures contain only single or several meta-atoms per period[25,26]. This approach gives a new perspective to design a simple and efficient anomalous reflective metasurfaces. Indeed, in this way, the restriction of the sub-wavelength cells/meta-atoms is highly released along with the advantages of reduced intrinsic loss and easy fabrication. With a suitable engineered arrangement of grooves, anomalous reflection and retroflection have been theoretically demonstrated[26]. However, imperfect pressure distribution composed of dominantly desired pattern with unexpected other orders of diffractions has been obtained. It stems from this result that only the propagating diffracting modes above the grating and only the basic mode inside the grooves were considered, and the interaction between the grooves within a period by which a lateral power flow exchanging along the surface can be fulfilled, was ignored.

In the present research, we provide a comprehensive concept of a perfect anomalous beam splitter based on a simple-structured acoustic meta-grating, capable of splitting a given acoustic wave with any incidence angle into two directions with arbitrary power flow partition. Compared to the previous metasurfaces with complex configurations and sub-wavelength units, only a sound hard surface containing two or several grooves per period is employed to construct the structure. By combining the general grating theory and a genetic optimization algorithm, a desired meta-grating can be easily created with feasible and practical design geometry. We evidence that the split directions and the corresponding power flow partition of the proposed splitter can be elegantly designed. Almost 100% energy conversion from a normal incident wave to a redirected reflection with 88° angle is validated. Experimental results also demonstrate that the power flow partition of the proposed splitter with 72° angle is accurately allocated to be 0:1, 1:1 and $\sqrt{5:3}$ respectively for the wave with frequency as higher as 8000Hz in air. Benefiting from the simplicity of the structure, in which extreme thin walls and narrow channels in cells/meta-atoms can be avoided, acoustic beam splitter for the higher working frequency can also be realized.

## Results

**The structure of meta-grating.** The structure we considered is schematically illustrated in Fig.1, which is the quasi-one-dimensional planar periodical sound hard surface having a pitch *a* in *x*-direction. In each period, there are *L* rectangular shaped grooves (*L=2* in Fig.1). The width and depth of the *l*th ($l=1,\cdots,N$) groove, and the distance between the *l*th and (*l+1*)th ones is denoted respectively by $t_l$, $d_l$ and $dx_l$. Because of the periodicity of the structure, an incident wave from the -*y* direction (surface normal) will be diffracted as $0^{th}$, $\pm 1^{st}$, and $\pm 2^{nd}$, $\cdots$, order diffractive components. The purpose of our design is to extinguish all the other propagation components except the ones in the desired directions.

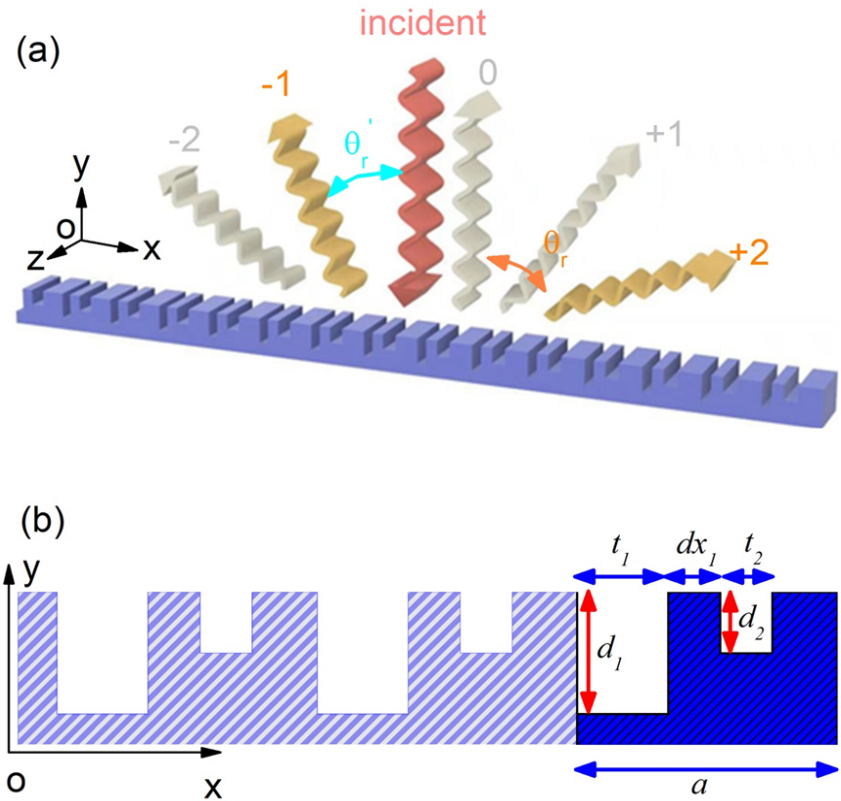

Fig.1 (a) Schematic illustration of the designed meta-grating, which consists of a quasi-one-dimensional planar period sound hard surface with etched rectangular grooves. The periodicity is along *x*-direction, and the period is *a*. An incident wave from the negative *y*-direction (surface normal) will be scattered as 0th, $\pm 1^{st}$, $\pm 2^{nd}$, $\cdots$, order diffractive components. (b) The cross section of the structure. The depth and width of the grooves in each period are denoted as $t_l$, $d_l$ ($l=1,2,...$), and the relative distance between the $l$th and ($l+1$)th grooves is denoted as $dx_l$.

**Grating theory for the structure.** We first show that, the diffractive field from such a structure can be rigorously solved by the general grating theory. According to the latter, when the periodic structure is illuminated by a planar pressure wave with incident angle $\theta_i$, the total field $p_s$ in the media above the surface can be expressed as[27,28]

$$p_s(x,y) = A_0^+ e^{-jk_0 x \sin q_i} e^{jk_0 y \cos q_i} + \sum_n A_n^- e^{-j(k_0 \sin q_i + G_n)x} e^{-jk_{yG_n} y} \tag{1}$$

where $k_0$ is the wave number of the media, $A_0^+$ is the amplitude of the incident wave and $A_n^-$ is the amplitude of the $n$th order harmonic component of the refractive wave. The symbol $G_n = n2\pi/a$ with $n=0,\pm 1,\pm 2,\cdots,\pm\infty$ is the $n$th reciprocal vector, and $k_{yG_n} = \sqrt{k_0^2 - (k_0 \sin\theta_i + G_n)^2}$ is the wavevector component in $y$-direction for the $n$th order harmonic mode. Notice that $k_{yG_n}$ will be real only when $k_0^2 > (k_0 \sin\theta_i + G_n)^2$. The infinite summation in the right-hand side of the equation includes only finite terms of propagation modes, and the total number and the direction of those propagating modes are controlled by the period $a$ of the structure.

To design an anomalous splitter which can reflect a wave with incident angle $\theta_i$ into reflected directions with angles $\theta_r$ and $\theta_r'$ ($\theta_r > \theta_r'$), we first can set the period $a$ by the formula $k_0 \sin\theta_r = k_0 \sin\theta_i + G_n$ to make sure the $m$th order refractive component being a propagating mode. This condition gives

$$a = m \frac{2p}{k_0(\sin q_r - \sin q_i)} \tag{2}$$

where $m$ takes a positive integer. However, because the condition $k_0^2 > (k_0 \sin\theta_i + G_n)^2$ can be satisfied by a number of $n$ under such setting, it means that a finite number of unwanted propagation modes with directions other than $\theta_r$ will co-exist in the diffractive field. Therefore, those unwanted modes have to be extinguished. We have found that this can be realized by adjusting the relative positions and the geometric parameters of the bottom-connected grooves.

The pressure wave field in the rectangular shaped grooves can be written as the superposition of the waveguide modes as

$$p_g^l = \sum_n H_{nl} \cos\alpha_{nl}(x-x_l)\left(e^{j\beta_{nl}y} + e^{-j\beta_{nl}(y+2d_l)}\right) \qquad (3)$$

where $p_g^l$ means the pressure wave in the $l$th groove, $H_{nl}$ means the amplitude of the $n$th order component of the waveguide mode, and $\alpha_{nl} = \dfrac{n\pi}{t_l}$ and $\beta_{nl} = \sqrt{k_0^2 - \alpha_{nl}^2}$ are respectively the $x$- and $y$-component of the wavevector for the $n$th order mode in the $l$th groove. By using Eqs.(1), (3) and the continuum condition for the pressure and surface-normal velocity field at the interface, we can get a linear equation set about $A_0^+$, $A_n^-$ and $H_{nl}$ as

$$Q_1 A_0^+ = Q_2 \begin{bmatrix} A^- \\ H \end{bmatrix} \qquad (4)$$

where $A^- = (A_1^-, \ldots, A_N^-)^T$ will be the $N$-order column vector when the summation of the harmonic modes in Eq.(1) is truncated with $N$ terms, and $H=(H_1,\cdots,H_l,\cdots H_L)^T$ with $H_l=(H_{1l},\cdots,H_{Ml})^T$, $(l=1,\cdots,L)$, will be the $LM$-order column vector when the summation of the waveguide modes in Eq.(3) is truncated with $M$ terms. A detailed derivation of Eq.(4) and the elements of the matrixes $Q_1$ and $Q_2$ can be found in the supplementary material.

By Eq.(4), the amplitudes $A_n^-$ ($n=1,\cdots,N$) can be solved under the given period $a$, the incident angle $\theta_i$, and the total number $L$ and the geometric parameters of the grooves. However, to obtain the suitable geometric parameters of the grooves for the desired metasurface, an optimization procedure with a searching target is needed. For instance, when we need to split the incident wave with $\theta_i=0$ into directions with $\pm\theta_r$ as the $\pm 1^{st}$ order refractive components, and require the power flow ratio between them to be $I_{-1}:I_{+1}=v:(1-v)$, we first can set $m=1$ in Eq.(2) to get $a=\lambda/\sin\theta_r$, where $\lambda$ is the working wavelength. Then, we start the optimization algorithm to search the structure with targeted function $f = \left|v - \left|\frac{A_1^-}{A_0^+}\right|^2\right| + \left|(1-v) - \left|\frac{A_{-1}^-}{A_0^+}\right|^2\right| \to 0$. In our calculation, the genetic algorithm (GA) is chosen as the optimization algorithm.

**Meta-grating with mirror symmetric splitting angle.** As a first demonstration, we choose to design the metasurfaces to reflect the normally incident waves into extreme directions with $\theta_r = -81°$ and $-88°$, respectively. For these purposes, we set first $a = \lambda/|\sin\theta_r|$ and then start the searching procedure for $I_{-1}:I_1 = 0:1$. To guarantee the convergence in the numerical calculation, the truncations for the summations in Eq. (1) and (3) are set as $N=101$ and $M=10$, respectively. In the optimization procedure, the depth of the grooves is limited in the range of $0 \leq d_l \leq 0.5\lambda$. To avoid ultrathin substructure, both of the width $t_l$ and the neighbor distance $dx_l$ of the grooves are limited to be greater than $0.05a$. By performing the searching procedure with 2 grooves per period (a detailled discussion about the necessary number of total grooves is provided in the supplementary material), we get the structure with $t_{1,2}=(0.500, 0.148)a$, $d_{1,2}=(0.399, 0.121)\lambda$ and $dx_1=0.111a$ for $\theta_r=-81°$, and the structure with $t_{1,2}=(0.085, 0.527)a$, $d_{1,2}=(0.419, 0.070)\lambda$ and $dx_1=0.101a$ for $\theta_r=-88°$. Both of the structures are obtained under the condition with $f<10^{-8}$ in the searching procedure. To verify the obtained structures, we perform a finite-element (FE) simulation based on COMSOL Multiphysics software. We show in Fig. 2(a) and (b) the Fast-Fourier-Transform (FFT) amplitudes of the obtained reflective fields at the distance away from $y=4a$. The diffractive pressure distributions (real part) in one period are shown also as insets in the same figure, respectively. It can be clearly seen from the figures that the diffractive field from the structures is almost completely in the desired directions. We also find that, for both cases, the amplitude of the field in and out of the grooves is definitely in the same order, which means that they are not resonating.

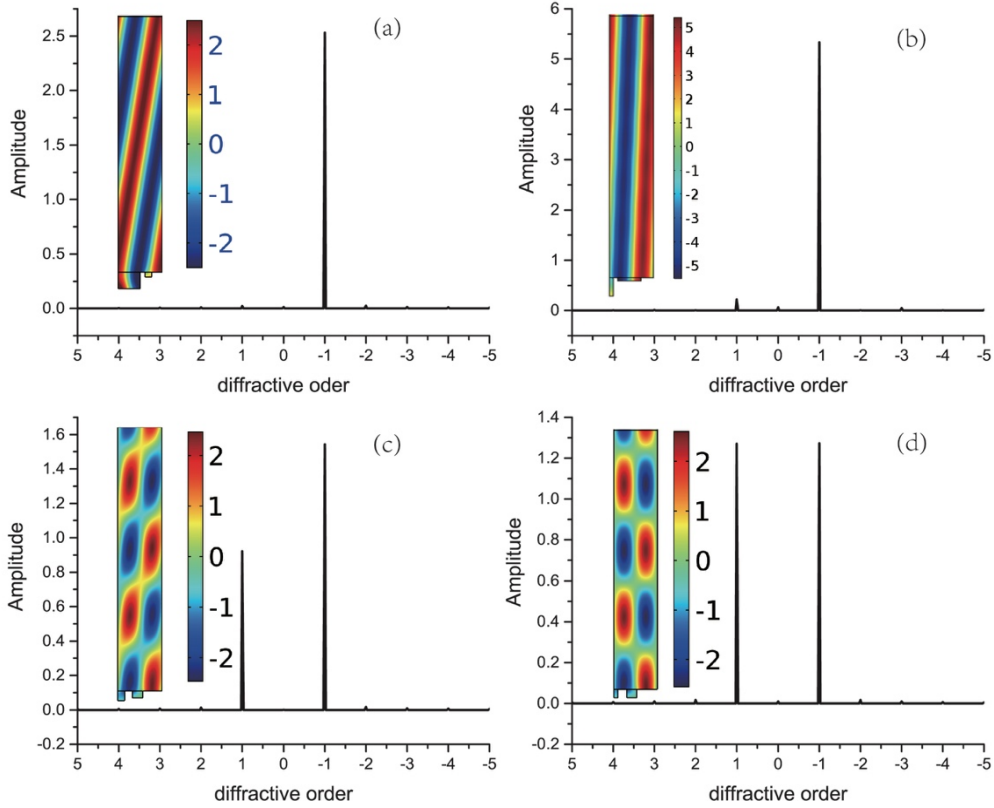

Fig.2 Pressure field distribution obtained by finite element method. Fast-Fourier-Transform amplitude of the reflecting pressure field from the meta-grating at $y=4a$ is present to check the efficiency of the reflector. The insets show the real part of the pressure field distribution (within the area $0<x<a$ and $y<6a$). (a) and (b) are for the reflectors with desired angles $\theta_r=-81°$ and $-88°$ respectively; (c) and (d) are for the splitters with desired angles $\theta_r=72°$. The power flow ratios for (c) and (d) are $I_{-1}:I_1=\sqrt{5:3}$ and $1:1$, respectively. For all of them, the incident plane wave is along $-y$ direction (not shown in the figure).

Similarly, to split the waves into the mirror symmetric directions with different power flows, we can design the structure by first setting $a=\lambda/|\sin\theta_r|$, and then setting $I_{-1}:I_1$ to be the desired value in the target function. Here, we choose to design two splitters with $\theta_r=\pm72°$, $I_{-1}:I_1=\sqrt{5:3}$ and $1:1$ (the corresponding amplitude ratio are $\left|\frac{A_{-1}^-}{A_1^-}\right|=5:3$ and $1:1$), respectively. Optimization procedure shows that a structure with $t_{1,2}=(0.159, 0.230)a$, $d_{1,2}=(0.229, 0.170)\lambda$ and $dx_1=0.176a$ can realize the former target, while a structure with $t_{1,2}=(0.094,0.229)a$, $d_{1,2}=(0.207,0.205)a$ and $dx_1=0.206a$ can realize the later one. We present in Fig. 2(c) and (d) the FFT results and the real part of the pressure field distributions of the reflecting waves. From Fig.2(c), we can obtain the amplitude ratio as $\left|\frac{A_{-1}^-}{A_1^-}\right|=\frac{1.5431}{0.9216}\approx 5:3$, and from Fig.2(d), we can obtain the amplitude ratio as $\left|\frac{A_{-1}^-}{A_1^-}\right|=\frac{1.2719}{1.2729}\approx 1:1$.

**Power flow redistribution by the grooves effect.** The perfect structure allows investigating and discussing the underlying physics of the metasurface. Without losing generalities, as example, we have chosen to study the metastructure with the reflected beam at $\theta_r=-81°$ [shown in Fig. 2(a)]. By inserting the structure parameters into Eq.(4), we can obtain respectively the total pressure $p_t$, the total velocity along $y$ direction $v_{yt}$, and the outgoing pressure waves $p_{out}$. With these values, the phase profile $P_s=angle(p_{out})$ and the $y$-directional power flow $I_{y0}=Re[p_t(v_{yt})^*]/2$ at the surface ($y=0$) are calculated. The values in one period in $x$ direction are shown in Fig. 3(a) and (b), respectively. From the figure we see that, the phase curve increases almost linearly with the slope $2\pi/a$ from $P_s=\theta_0$ to $\theta_0+2\pi$ with $\theta_0$ as an initial phase. This indicates the linear gradient-phase profile along the surface, implying that the basic requirement in the gradient-phase metasurfaces is satisfied. As for the $I_{y0}$ curve, it is fluctuated around zero in the whole period. In details, it rapidly fluctuates with very small amplitude in the area without grooves, while in the area with connected grooves, the curve symmetrically turns from positive to negative values with large amplitude (for groove 1) or vice versa (for groove 2). This means that the power flow is conserved in the period but redistributed along the surface by the grooves effect.

To further understand this phenomenon, we have checked the local intensity vector distribution $\vec{I}_t=I_x\vec{x}+I_y\vec{y}$ in and above the grooves, where $I_i$, ($i=x,y$) can be calculated by $I_i=Re[p_t(v_{it})^*]/2$ from the obtained field, and $v_{it}$ is the $i$-directional component of the total velocity field. The result within $y<0.8a$ in $y$ direction and one period in

*x*-direction is shown in Fig. 3(c), in which the amplitude of the local intensity vectors is shown by the length of the arrows and its direction is shown by the arrows. It can be found from the figure that, except some distortions near the interface $y\approx0$, the arrows show a regular pattern in a certain distance away from the surface. This means that there is a lateral energy exchanging along the surface. In terms of physical mechanism, this lateral energy exchanging is fulfilled by the high-order evanescent harmonic modes reshaped by the grooves. More interestingly, unlike the narrower groove in the right-hand side in the Fig. 3 (c), which distorts the vectors intensity only in a small region in lateral direction, the wider groove in the left-hand side makes the energy exchanging happens in a wide lateral area. This effect is finished by a vortex-shaped flow inside the groove. It can be seen that, by this vortex, the desired local intensity vector pattern above the surface is fitted by arrows leading the flow in and then out of the groove from the left-hand to right-hand side. Notice that to form such a vortex in the groove, higher orders real (rather than evanescent) waveguide modes than the *0*th order is needed.

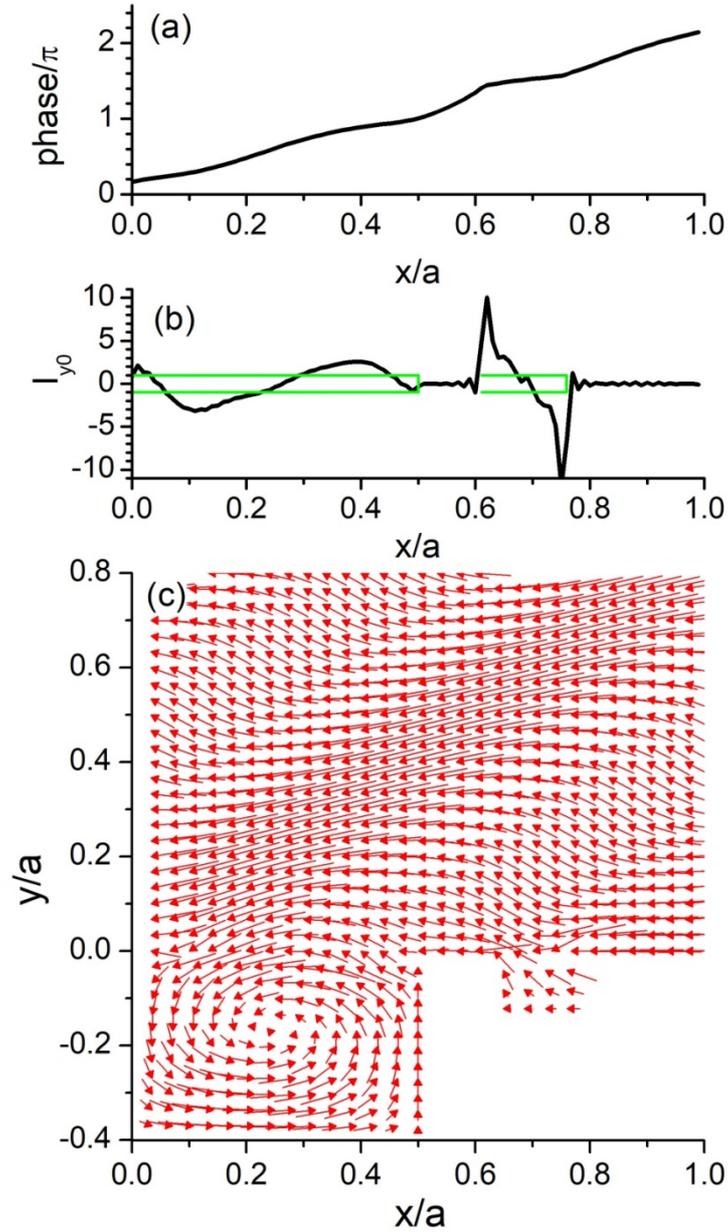

Fig.3 (a) The phase profile of the diffractive pressure wave, (b) the *y*-directional power flow of the diffractive wave at *y=0* in one period in *x* direction (the positions where the grooves are connected are shown by green boxes), and (c) local intensity vector distribution for the total field in the structure with $\theta_r=-81°$. The incident plane wave is along –*y* direction.

**Experimental measurements.** To verify the numerical result, we select three structures under $\theta_r=\pm72°$ with *I₋₁:I₁=0:1, 1:1* and $\sqrt{5:3}$ for experiment. Geometrical parameters for the structures with $I_{-1}:I_1=\sqrt{5:3}$ and 1:1 have been listed above [as the same as the ones for Fig.3(c) and (d)]. Parameters for the one with *I₋₁:I₁=0:1* are obtained by the optimization algorithm. It gives $t_{1,2}=(0.182, 0.480)a$, $d_{1,2}=(0.129, 0.361)\lambda$ and $dx_1=0.071a$. In our experiment, we choose the air as working medium and the working frequency as $f=8000Hz$ ($\lambda=42.88mm$). Under this frequency, the narrowest groove and the thinnest wall in all three samples is about 4mm (for the structure with $I_{-1}:I_1=1:1$) and 3.2mm (for the structure with $I_{-1}:I_1=0:1$), respectively. This means that the additional effects of friction and wall deformation caused by the narrow channel and thin wall can be neglected. This frequency is much higher than the one (usually about 3000Hz) used for structures suggested in previous literature. We point out that, because narrow channels and thin walls (compared to the working wavelength) have to be used in the structures suggested in previous works, it is very difficult to push their working frequency into the region as high as 8000Hz.

We show in Fig. 4(a), (b) and (c) the pressure field distribution for structures with $I_{-1}:I_1=0:1$, 1:1 and $\sqrt{5:3}$, respectively. The left panels present the simulated results and the right panels present the corresponding experimental ones measured in the areas marked by red boxes in the left panel. A Good agreement between the simulation and experimental results are obtained. Notice that the incident beam is not shown for clear eyesight.

To clearly show the power flow partition in Fig. 4(b) and (c), measured value of the pressure field along $y=115mm$ away from the surface (marked by dash line in the right panels of the figures) is extracted out and plotted in Figs. 4(d) and (e), respectively. In the latter, the lateral axes are defined as $D(\theta)=Re(p_\theta/p_{max})$ with $\theta$ as the angle between the $x$-axis and the line from the central of the incident beam to the measured point. The amplitude ratio around 1:1 in Fig. 4(d) and 5:3 in Fig. 4(e) can be verified.

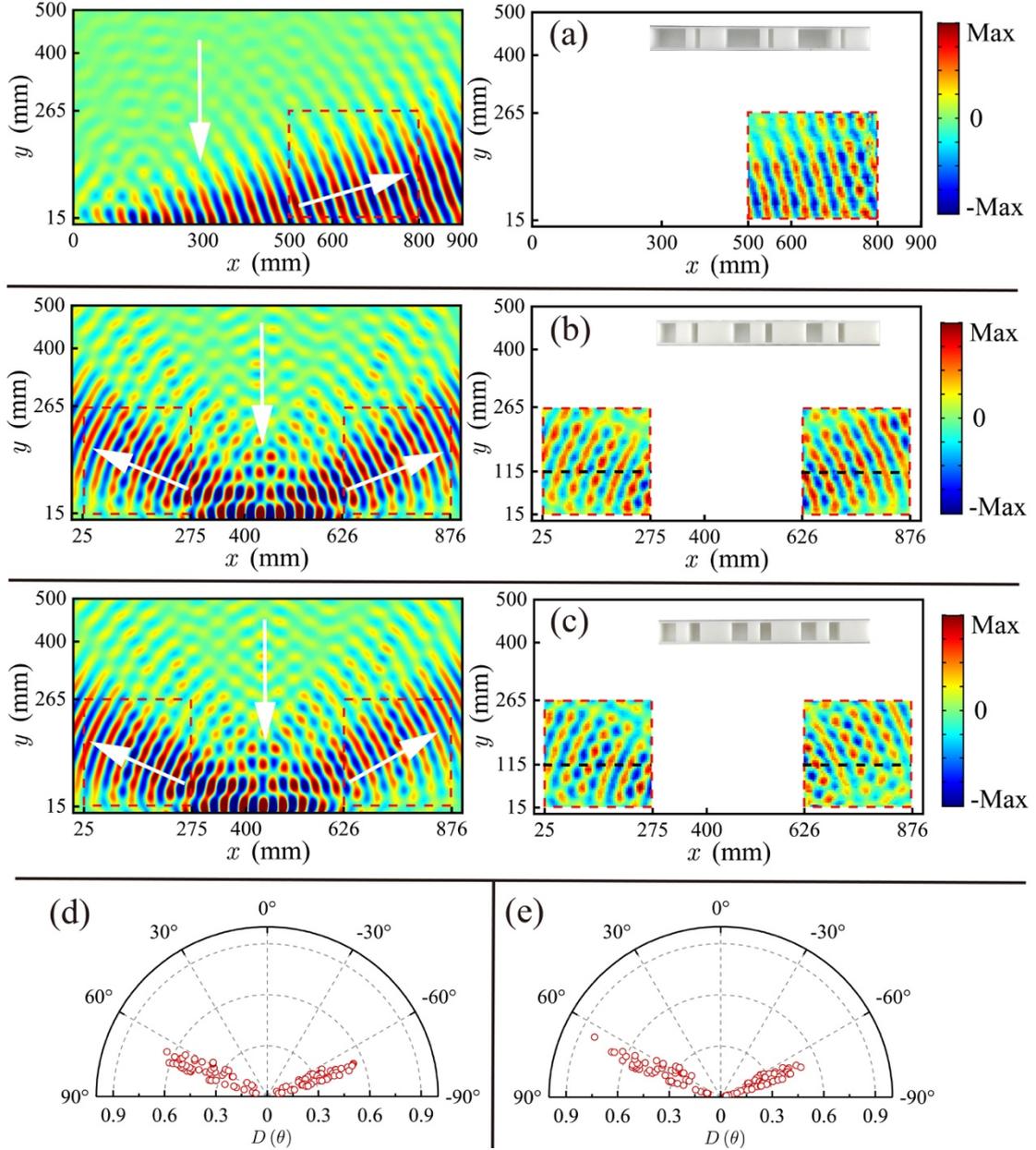

Fig.4 Pressure wave distribution of the diffraction wave from the structure for $\theta_r=\pm72°$. The structures contains totally 20 periods. (a), (b) and (c) are for structures with desired power flow ratios $I_{-1}:I_1=0:1$, $1:1$ and $\sqrt{5}:3$, respectively. The left panels are results simulated by finite element method, and the right panels are the measured data in the areas marked by the red boxes in the corresponding left panels. The insets in the right panels are the photo of the samples (three periods are shown). For all of them, the incident wave along $-y$ direction is the beam with rectangle-shaped amplitude distribution in $x$-direction. (d) and (e) are the measured data extracted along the dash lines marked in (b) and (c), respectively. The lateral axes in (d) and (e) are defined

as $D(\theta)=Re(p_\theta/p_{max})$ with θ as the angle between the *x*-axis and the line from the central of the incident beam to the measured point.

**Meta-grating with different splitting angles.** Finally, we would like to show that perfect splitters can also be designed to steer the reflective waves into two different directions with arbitrary power flow partition. In our best knowledge, such kind of structures and effect has not been reported yet. As an example, we chose to design a splitter that can steer the normally incident wave into directions with $\theta_r=81°$ and $q_r^{'}=-a\sin(\sin q_r/2)=-28.4°$ with equally distributed power flow. To design such a structure, we first need to set the period by letting *m=2* in Eq.(2), which gives $a=2\lambda/|\sin\theta_r|$. For such a structure, there will be totally 5 propagating modes in the refractive field. The refractive angles for the $\pm 2^{nd}$ and $\pm 1^{st}$ order components are $\pm\theta_r$ and $\pm q_r^{'}=\pm a\sin(\sin q_r/2)$, respectively. With the period *a*, we search the optimization structure by minimizing the target function $f=\left|\cos q_r\left|\frac{A_{-1}^-}{A_0^+}\right|^2-0.5\right|+\left|\cos q_r^{'}\left|\frac{A_{-2}^-}{A_0^+}\right|^2-0.5\right|$. Under these setting, a structure with 4 grooves per period having the detailed parameters: $t_1\sim t_4$=(0.101, 0.107, 0.101, 0.104)*a*, $d_1\sim d_4$=(0.186, 0.321, 0.126, 0.254)*λ*, and $dx_1\sim dx_3$=(0.105, 0.101, 0.101)*a* is obtained. To intuitively show the effect and to verify the numerical result, we have constructed a finite structure with 15 periods, and illuminate it from *–y* direction by an equal-amplitude beam. The FE result of the diffractive pressure field distribution (real part) and the corresponding measured data are shown in Fig. 5(a) and (b), respectively. It can be seen from the figure that, the beam splitting effect is perfect, and the agreement between the numerical and experiment result is good. It can be found that, the measured pressure field in the left marked region is weaker than the corresponding simulation result. This is due to the influence by the second reflection from the speaker array.

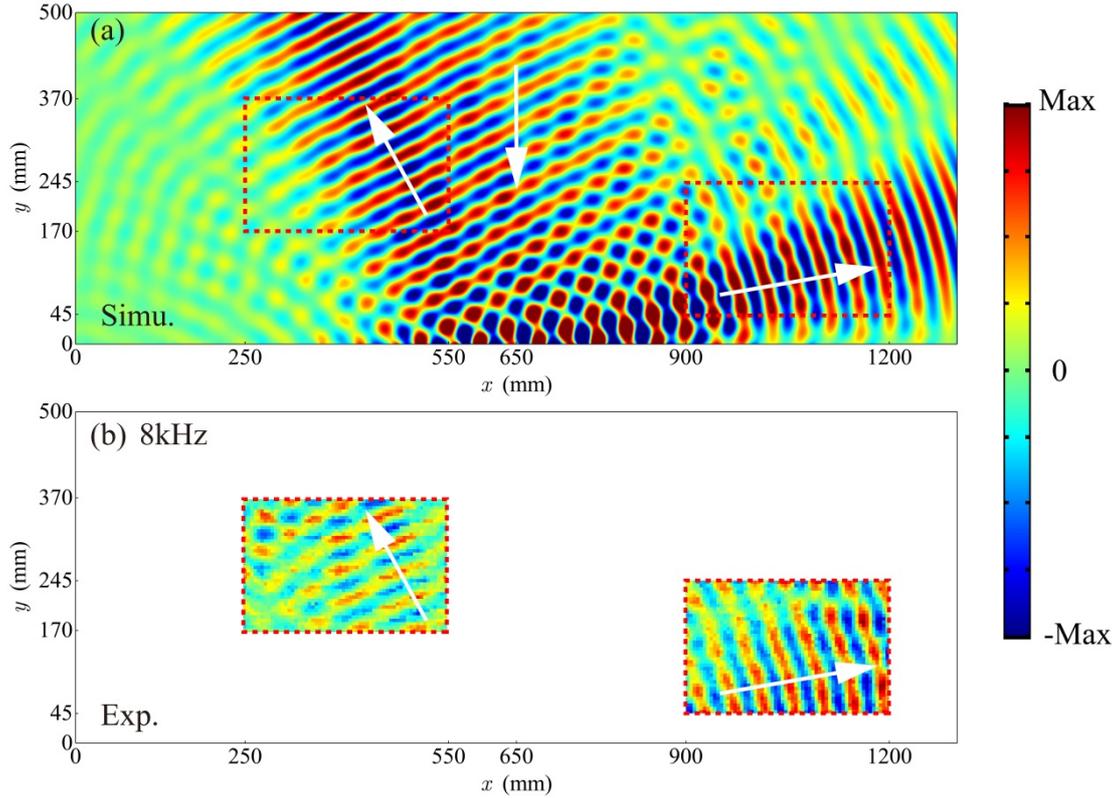

Fig.5 (a) Diffractive pressure field distribution (real part) of the structure illuminated by an equal-amplitude beam with frequency $f$=8000Hz along $-y$ direction. The structure contains totally 15 periods. The width of the incident beam is 410mm in $x$ direction. White arrows show schematically the directions of the incident and the deflecting beams. (b) Measured data in the areas marked by red boxes in (a).

## Discussion

In conclusion, we developed a numerical approach to solve the diffraction problem for the sound hard surface with periodically etched grooves. By using this approach and an optimization algorithm, meta-gratings which can split the incident wave into desired directions with arbitrary power flow partition are theoretically and experimentally demonstrated. The predicted structures are verified by the finite element simulation and experiment. The developed method can be easily extended to design metasurfaces for other purposes. In contrast to the classical metasurfaces which usually having many arranged sub-wavelength meta-atoms, the structures provided and designed by our method have only two or four straight-walled grooves per period. Because thin walls and narrow channels in the meta-atoms design are avoided, the additional friction and wall deformation can be suppressed. This means that the working frequency of our concept can be much higher than the ones suggested in previous literature.

## Method

*Numerical simulation.* The full wave simulations based on finite element analysis are performed using COMSOL Multiphysics Pressure Acoustics module. For the left panels in Fig.2 (a)-(d), plane wave along –$y$ direction is chosen as incident wave. The Perfectly Matched Layers (PML) with thickness $2a$ are added at the top regions (not shown in the figure) to reduce the reflection on the boundaries. Floquet periodic boundary condition is added on the left and right boundaries. For the left panels of Fig.4 (a)-(c), the equal-amplitude beam with finite width in $x$ direction is chosen as incident wave. The plane wave radiation (PWR) boundary condition on the top, left and right boundaries are used. For Fig. 5(a), the equal-amplitude beam is used as the source, PML with thickness $2a$ is added on the top region (not shown in the figure), and PWRs are used on the left and right boundaries.

*Experimental apparatus.* The samples with 20 periods are fabricated using the stereo lithography apparatus (SLA) with photosensitive resin. The molding thickness of each layer during printing is 0.1mm. Organic Glass plates are added on the top and bottom of the samples to form a two-dimensional wave guide for measurement. Foams are distributed on sides of the waveguide to absorb sound wave with frequency above 3000Hz.

For Fig. 4(a), an array of 22 loudspeakers is used as a source, and for Fig. 4(b) and (c), an array of 7 loudspeakers is used for the same purpose. The loudspeakers are controlled by the generator with the same amplitude at *f=8000*Hz. Signals are collected by Brüel & Kjær Data Acquisition Hardware (LAN-XI, type 3160-A-042). The probe is a 1/8-inch microphone (Brüel & Kjær type-2670).

The field in the areas marked by red boxes [see the left panels in Fig. 4(a)-(c)] is scanned using a moving probe with a step of 5 mm. The measured area is 250mm×250mm for Fig. 4(a) and 300mm×200mm for Fig. 4(b) and (c), respectively. For Fig.5(b), an array of 11 loudspeakers with 410mm span is used as the source. The working frequency is $f$=8000Hz, and both of the measuring areas are 300mm×200mm.

# Acknowledgements

This work is supported by the National Natural Science Foundation of China (Grant No: 11274121).

*phzlhou@scut.edu.cn
*yongli@tongji.edu.cn
*badreddine.assouar@univ-lorraine.fr